\documentclass{article} %

\usepackage[final]{colm2026_conference}

\usepackage{microtype}
\usepackage{hyperref}
\usepackage{url}
\usepackage{amsmath}
\usepackage{booktabs}
\usepackage{graphicx}
\usepackage{xcolor}
\usepackage{multirow}
\usepackage{xspace}
\usepackage{wrapfig}

\usepackage{lineno}

\definecolor{darkblue}{rgb}{0, 0, 0.5}
\hypersetup{colorlinks=true, citecolor=darkblue, linkcolor=darkblue, urlcolor=darkblue}

\newcommand{\MODEL}{{Olmo~3}\xspace}

\title{The Hidden Cost of Thinking: Energy Use and Environmental Impact of LMs Beyond Pretraining}

\author{
Jacob Morrison \\
University of Washington \\
Allen Institute for AI
\And
Noah A. Smith \\
University of Washington \\
Allen Institute for AI
\And
Emma Strubell \\
Carnegie Mellon University
}

\begin{document}

\ifcolmsubmission
\linenumbers
\fi

\maketitle

\begin{abstract}
Modern language model development extends far beyond pretraining, yet environmental reporting remains narrowly focused on the cost of training a single final model. In this work, we provide the first detailed breakdown of the environmental impact of a full model development pipeline, from pretraining through supervised fine-tuning, preference optimization, and reinforcement learning, for \MODEL, a family of 7 billion and 32 billion parameter models in both instruction-following and reasoning variants. We find that \textbf{reasoning models are 17$\times$ more expensive to post-train} than their instruction-tuned counterparts in terms of datacenter energy, driven by reinforcement learning rollout generation. Development costs (including experimentation, failed runs, and ablations) account for \textbf{82.2\%} of total compute, a roughly $65\%$ increase over the $\sim$50\% reported for pretraining-focused pipelines in prior work. In total, we estimate our model development process consumed \textbf{${\sim}$12.3~GWh} of datacenter energy, emitted \textbf{4,251~tCO$_2$eq}, and consumed \textbf{15,887~kL} of water, with water consumption driven entirely by power generation infrastructure rather than data center cooling. These costs, which are almost entirely unreported by model developers, are growing rapidly as post-training pipelines become more complex, and must be accounted for in environmental reporting standards and by the research community developing solutions to reduce AI's environmental impact.
\end{abstract}

\section{Introduction}
\label{sec:intro}

The environmental cost of building AI systems has grown dramatically alongside their capabilities~\citep{schwartz2019greenai,strubell2019energypolicyconsiderationsdeep,wu2022sustainableaienvironmentalimplications}. As model developers build increasingly powerful systems, the energy consumed, carbon emitted, and water used throughout development continue to rise. Yet the way we \emph{report} these costs has not kept pace with how models are actually built. Most environmental reporting focuses narrowly on the cost of pretraining a single final model~\citep{grattafiori2024llama3herdmodels,gemmateam2024gemmaopenmodelsbased}, leaving the majority of the development process invisible.

Modern language model development extends far beyond pretraining. A state-of-the-art model pipeline now involves pretraining, mid-training on curated data, long-context extension, large-scale synthetic data generation, supervised fine-tuning (SFT), preference optimization (e.g. DPO; \cite{rafailov2024directpreferenceoptimizationlanguage}), and reinforcement learning (RL), each stage requiring its own extensive experimentation cycle before a final configuration is selected. The rise of reasoning models, agentic workflows, and tool use training has made post-training particularly compute-intensive: reinforcement learning with verifiable rewards (RLVR) requires generating extremely large numbers of long rollouts.

In this paper, we provide the first detailed accounting of the environmental impact of this full pipeline. We analyze the development of the \MODEL family of 7 billion and 32 billion parameter language models, each released in instruction-following and reasoning (``thinking'') variants. Our methodology follows prior work~\citep{morrison2025holisticallyevaluatingenvironmentalimpact,luccioni2022estimatingcarbonfootprintbloom}, with one improvement: we account for non-GPU server and IT infrastructure power using datacenter overhead estimates from \citet{epochai2025gpu}. Our primary contribution is not methodological but empirical: to the best of our knowledge, we are the first to measure and report the environmental cost of each post-training stage, and the first to compare reasoning and instruction-tuned variants. Building on \citet{morrison2025holisticallyevaluatingenvironmentalimpact}, where we previously reported development costs of approximately 50\% of total compute, we find that the landscape has shifted substantially.

Our main findings and contributions are:
\begin{itemize}
    \item We provide the \textbf{first detailed breakdown of the environmental cost of post-training} stages (SFT, DPO, RL) and synthetic data generation, enabling comparison across stages and model variants.
    \item We show that \textbf{reasoning models are far more expensive to post-train} than instruction-tuned models: \MODEL 32B Think's post-training consumes $17\times$ more datacenter energy than 32B Instruct, driven by RL rollout generation, which accounts for 87\% of Think-specific post-training energy.
    \item We report that \textbf{development costs have grown to 82.2\%} of total compute, up from $\sim$50\% reported by \citet{morrison2025holisticallyevaluatingenvironmentalimpact}, a finding corroborated by concurrent financial analyses at frontier-scale organizations~\citep{denain2026rnd}.
\end{itemize}

In total, developing the \MODEL series emitted an estimated 4,251 tCO$_2$eq, equivalent to approximately 847 US homes' electricity for one year, and consumed an estimated 15,887~kL of water. We additionally analyze water consumption and find that it is driven primarily by power generation infrastructure rather than AI workloads, with implications for how water costs should be attributed and addressed. We release our detailed per-stage measurements to support future work on environmental accounting of AI systems. As post-training pipelines grow in complexity and cost, these largely unreported impacts will only increase. Accurate environmental accounting that captures the full development lifecycle is essential for informed regulatory guidelines and for the research community to develop effective strategies to reduce the environmental footprint of AI.

\begin{figure}[t]
\begin{center}
\includegraphics[width=\linewidth]{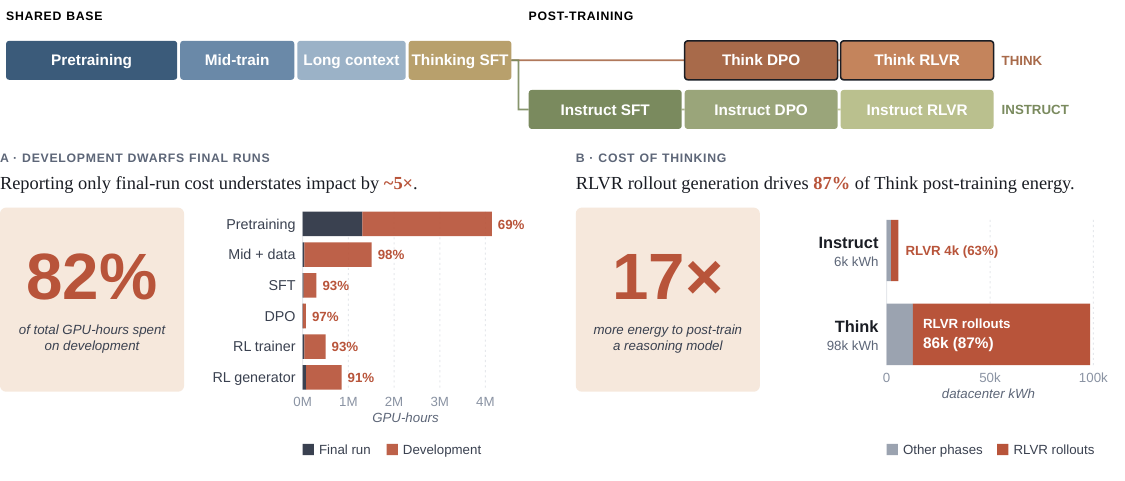}
\end{center}
\vspace{-2em}
\caption{Building \MODEL cost 8.34M GPU-hours across pretraining, mid-training, long-context, and post-training, with all variants sharing a base before branching into Think and Instruct (top). 82\% of that compute went to development (e.g. hyperparameter searches, ablations, and failed runs) so reporting only final-run cost understates impact by ${\sim}5\times$ (bottom left). Within 32B post-training, Think uses $17\times$ more datacenter energy than Instruct, with 87\% of that energy spent on RLVR rollout generation (bottom right).}
\label{fig:pipeline}
\end{figure}

\section{Related work}
\label{sec:related}

\paragraph{Reporting the environmental impact of training.} While most model developers do not report environmental impact data, a growing number have included partial estimates. \citet{luccioni2022estimatingcarbonfootprintbloom} provided the most comprehensive analysis to date for the BLOOM model, reporting embodied emissions from hardware manufacturing, electricity consumption during training, and idle cluster power. \citet{dodge2022measuringcarbonintensityai} measured electricity consumption and carbon emissions with granular timesteps and region-specific carbon intensity, but did not measure water consumption or development costs. The Llama model reports~\citep{touvron2023llamaopenefficientfoundation,touvron2023llama2openfoundation,grattafiori2024llama3herdmodels} included electricity consumption and carbon emissions for their final training runs, though their approach to carbon intensity varied across releases and all three assumed 100\% GPU power draw throughout training. \citet{gemmateam2024gemmaopenmodelsbased} reported only a single aggregate emissions figure for pretraining, not broken down by model or stage. The Olmo reports~\citep{groeneveld2024olmoacceleratingsciencelanguage,olmo20252olmo2furious} documented electricity consumption per model with region-specific carbon intensity, and Olmo~2 additionally estimated water consumption. Across all of these works, reporting is limited to the \emph{final training run} of released models.

\paragraph{The hidden costs: development, post-training, and inference.} \citet{morrison2025holisticallyevaluatingenvironmentalimpact} provided the first public estimates of the environmental impact of LLM \emph{development} (the hyperparameter searches, failed runs, and ablations that precede a final training run) and found that development costs were approximately equal to the cost of training the final models. However, that analysis focused on pretraining and did not break down post-training costs. To the best of our knowledge, no prior work has reported a detailed breakdown of the environmental cost of post-training stages such as supervised fine-tuning (SFT), direct preference optimization (DPO), or reinforcement learning (RL). This gap is increasingly consequential: as reasoning models become widespread, post-training pipelines grow more complex and computationally demanding, yet these costs remain unreported. Concurrently with this work, \citet{denain2026rnd} estimated from financial disclosures that final training runs account for only 10--23\% of total R\&D compute spending at three major AI companies, providing independent evidence that development costs dominate. Separately, \citet{Luccioni_2024} estimated the energy cost of \emph{deploying} AI systems, highlighting that inference costs can rival or exceed training costs at scale.

\paragraph{Water consumption and embodied emissions.} Comparatively little transparency has been provided on non-carbon environmental impacts. \citet{li2025makingaithirstyuncovering} estimated the water consumption of closed models such as GPT-3, but these estimates rely on speculation about training location and energy consumption. \citet{morrison2025holisticallyevaluatingenvironmentalimpact} measured water consumption for a series of open models, finding it to be substantial even with an efficient data center, and emphasized the role of both cooling infrastructure and power generation methods in determining total water use. Embodied emissions from hardware manufacturing remain similarly opaque: \citet{wu2022sustainableaienvironmentalimplications} and \citet{luccioni2022estimatingcarbonfootprintbloom} highlighted that researchers are forced to rely on rough estimates, and the situation has not improved, as GPU manufacturers still do not disclose the environmental impact of their products. In this work, we extend prior analyses by providing the first detailed accounting of post-training environmental costs, comparing reasoning and instruction-tuned model variants, and analyzing the factors that drive water consumption.

\section{Methodology}
\label{sec:methodology}

\subsection{Operational impacts}
\label{sec:operational}

Following prior work~\citep{schwartz2019greenai,luccioni2022estimatingcarbonfootprintbloom,morrison2025holisticallyevaluatingenvironmentalimpact}, we estimate carbon emissions as $\text{CO}_2\text{e} = P_{\text{DC}} \cdot CI$ and water consumption as $\text{Water} = P_{\text{DC}} \cdot (WUE_{\text{onsite}} + WUE_{\text{offsite}})$~\citep{li2025makingaithirstyuncovering}. We measure per-GPU power at sub-second intervals and estimate total datacenter power as $P_{\text{DC}} = P_{\text{GPU}} \times 1.74 \times PUE$, where the $1.74\times$ factor accounts for non-GPU server and IT infrastructure overhead~\citep{epochai2025gpu}, improving on prior work that used only GPU power $\times$ PUE. Table~\ref{tab:methodology-params} summarizes all parameters used in our calculations. We additionally estimate embodied carbon and water by amortizing per-GPU manufacturing impacts over a 4-year lifespan~\citep{luccioni2022estimatingcarbonfootprintbloom,morrison2025holisticallyevaluatingenvironmentalimpact}; GPU manufacturers do not disclose this data, and these estimates remain rough approximations.

\begin{table}[t]
\begin{center}
\small
\begin{tabular}{llr}
\toprule
\textbf{Parameter} & \textbf{Description} & \textbf{Value} \\
\midrule
\multicolumn{3}{l}{\textit{Operational}} \\
\midrule
Server/IT overhead & Non-GPU power factor~\citep{epochai2025gpu} & $1.74\times$ \\
$PUE$ & Power usage effectiveness & 1.2 \\
$CI$ & Carbon intensity & 0.332 kg CO$_2$/kWh \\
$WUE_{\text{onsite}}$ & DC cooling (closed-loop) & 0 L/kWh \\
$WUE_{\text{offsite}}$ & Power generation~\citep{reig2020guidance} & 1.29 L/kWh \\
\midrule
\multicolumn{3}{l}{\textit{Embodied~\citep{luccioni2022estimatingcarbonfootprintbloom,morrison2025holisticallyevaluatingenvironmentalimpact}}} \\
\midrule
Carbon per GPU hr & Amortized over 4-year lifespan & 0.013 kg CO$_2$eq \\
Water per GPU hr & Amortized over 4-year lifespan & 0.003 L \\
\bottomrule
\end{tabular}
\end{center}
\caption{Parameters used for environmental impact calculations. All values are for our primary H100 training cluster. The combined datacenter overhead is $1.74 \times 1.2 = 2.088\times$ GPU power.}
\label{tab:methodology-params}
\end{table}

\subsection{Models and training pipeline}
\label{sec:models}

We analyze the environmental impact of developing the \MODEL family of language models, consisting of a 7 billion and a 32 billion parameter dense transformer, each released in two variants: an instruction-following model (\MODEL Instruct) and a reasoning model (\MODEL Think). Full architectural and training details are available in the \MODEL technical report~\citep{olmo2025olmo3}.

\paragraph{Training pipeline.} The \MODEL pipeline, illustrated in Figure~\ref{fig:pipeline}, consists of a shared base model that branches into variant-specific post-training. The shared base proceeds through pretraining, mid-training, long-context extension, and thinking SFT. From this shared base, the Instruct variant goes through instruction SFT, DPO, and RLVR, while the Think variant goes through thinking DPO and RLVR. Notably, the thinking SFT stage is shared between both variants, as it improved math and code performance for the Instruct model without increasing output lengths.

\paragraph{Reinforcement learning with verifiable rewards.} RLVR consists of two distinct compute workloads: a \emph{generator} that produces rollouts via autoregressive inference, and a \emph{trainer} that computes policy gradient updates on those rollouts. Generation dominates both wall-clock time and energy consumption, as we analyze in detail in Section~\ref{sec:cost-of-thinking}.

\paragraph{Data generation.} We also generated synthetic training data on the Frontier supercomputer using MI250X GPUs (${\sim}$4.9M GPU hours), producing data for SFT, DPO, and RLVR prompt filtering (discarding prompts where the model's pass rate was too high to provide useful learning signal).

\subsection{Hardware and data centers}
\label{sec:hardware}

All model training and development was conducted on the Jupiter cluster, hosted by Cirrascale and located in Texas, consisting of NVIDIA H100-80GB GPUs in standard HGX server configurations with 8 GPUs per node and high-speed interconnect between nodes. Data generation was performed on the Frontier supercomputer using AMD MI250X GPUs. The datacenter efficiency values used in our calculations ($PUE = 1.2$, $CI = 0.332$~kg CO$_2$/kWh, $WUE_{\text{offsite}} = 1.29$~L/kWh) were obtained directly from Cirrascale and its power provider, Austin Energy. See Section~\ref{sec:operational} for details on how these values are used in our calculations.

\section{Results}
\label{sec:results}

\subsection{Building our models}
\label{sec:building}

In this section we report a full accounting of the environmental impact of developing and training the \MODEL series, from data generation through development and final training runs. We follow the methodology outlined in Section~\ref{sec:methodology}.

Table~\ref{tab:final-runs} presents the per-stage cost breakdown for our final training runs. As expected, pretraining dominates in absolute terms: the 32B pretraining run alone accounts for ${\sim}$1.05M GPU hours and an estimated 1,423 MWh of datacenter energy. However, the post-training stages, while smaller individually, are collectively non-trivial, particularly for the Think variants. The total cost of our final training runs across both model sizes and all variants was ${\sim}$1.49M GPU hours, consuming an estimated 1,947 MWh of datacenter energy and emitting 647~tCO$_2$eq.

\begin{table}[t]
\begin{center}
\small
\begin{tabular}{lrrrr}
\toprule
\textbf{Component} & \textbf{GPU Hrs} & \textbf{DC kWh} & \textbf{CO$_2$ (kg)} & \textbf{Water (kL)} \\
\midrule
7B shared base & 246,513 & 315,366 & 104,701 & 407 \\
32B shared base & 1,120,912 & 1,514,671 & 502,871 & 1,954 \\
\midrule
7B Think post-train & 10,046 & 8,524 & 2,830 & 11 \\
7B Instruct post-train & 4,169 & 4,646 & 1,542 & 6 \\
32B Think post-train & 96,561 & 98,464 & 32,690 & 127 \\
32B Instruct post-train & 7,128 & 5,659 & 1,878 & 7 \\
\midrule
\textbf{Total (final runs)} & \textbf{1,485,329} & \textbf{1,947,330} & \textbf{646,512} & \textbf{2,512} \\
\bottomrule
\end{tabular}
\end{center}
\caption{Summary of final training run costs for \MODEL 7B and 32B. The shared base model (pretraining through thinking SFT) is used by both Think and Instruct variants. Post-training for the 32B Think model consumes $17\times$ more datacenter energy than 32B Instruct, driven by RLVR generation. DC kWh is estimated total datacenter energy (GPU kWh $\times$ 1.74 server/IT overhead $\times$ 1.2 PUE). CO$_2$ uses $CI = 0.332$~kg/kWh; water assumes $WUE_{\text{offsite}} = 1.29$~L/kWh with $WUE_{\text{onsite}} = 0$ (closed-loop cooling). See Table~\ref{tab:detailed-appendix} for the full per-stage breakdown.}
\label{tab:final-runs}
\end{table}

\paragraph{Data generation.} We used ${\sim}$4.9M MI250X GPU hours on the Frontier supercomputer to generate synthetic training data (see Section~\ref{sec:models}). Each MI250X chip has two Graphics Compute Dies (GCDs), which we treat as separate logical GPUs at 200W average power per GCD, compared to ${\sim}$600W per H100 for equivalent workloads. The resulting impact is an estimated 675~tCO$_2$eq and 2,621~kL of water. MI250X GCD hours are not directly comparable to H100 GPU hours in compute capability, so we report them separately.

\paragraph{Putting it in perspective.} Table~\ref{tab:environmental-summary} summarizes the total environmental impact of the \MODEL series across final training runs, development, and data generation. In total, we estimate that developing these models emitted approximately 4,251 tCO$_2$eq and consumed approximately 15,887~kL of water. Using the U.S.\ EPA's Greenhouse Gas Equivalencies Calculator, the carbon emissions are equivalent to approximately 847 US homes' electricity for one year, or about 56 tanker trucks' worth of gasoline burned. The water consumed is equivalent to approximately 140 years of water consumption by the average person in the United States. For comparison, \citet{morrison2025holisticallyevaluatingenvironmentalimpact} reported 493 tCO$_2$eq and 2,769~kL of water for a series of models up to 13B parameters with a pretraining-focused pipeline; our totals are roughly $8\times$ and $6\times$ larger, respectively, reflecting both larger models and a substantially more complex development pipeline. We discuss the drivers of water consumption in Section~\ref{sec:water}.

\begin{table}[t]
\begin{center}
\small
\begin{tabular}{lrrr}
\toprule
\textbf{Category} & \textbf{CO$_2$ (tCO$_2$eq)} & \textbf{Water (kL)} & \textbf{GPU Hours} \\
\midrule
Final training runs & 647 & 2,512 & 1,485,329 \\
Development & 2,757 & 10,714 & 6,849,712 \\
Data generation & 675 & 2,621 & 4,866,064 \\
Embodied (hardware) & 172 & 40 & -- \\
\midrule
\textbf{Total} & \textbf{$\sim$4,251} & \textbf{$\sim$15,887} & \textbf{$\sim$13.2M} \\
\bottomrule
\end{tabular}
\end{center}
\caption{Total environmental impact of developing the \MODEL series, including final training runs, development, data generation, and embodied hardware manufacturing. Carbon emissions of $\sim$4,251 tCO$_2$eq are equivalent to approximately 847 US homes' electricity for one year. Water consumption of $\sim$15,887 kL is equivalent to approximately 140 years of water use by the average US person. Embodied estimates are amortized over a 4-year GPU lifespan following \citet{luccioni2022estimatingcarbonfootprintbloom,morrison2025holisticallyevaluatingenvironmentalimpact}. All operational water consumption comes from power generation ($WUE_{\text{onsite}} = 0$ due to closed-loop cooling).}
\label{tab:environmental-summary}
\end{table}

\subsection{The cost of thinking}
\label{sec:cost-of-thinking}

A key contribution of this work is the first detailed comparison of the environmental cost of training reasoning models versus standard instruction-tuned models. As shown in Table~\ref{tab:final-runs}, the Think and Instruct variants share a common base model but diverge substantially in post-training cost. This cost is justified by substantial performance improvements that earlier post-training stages cannot achieve alone: compared to DPO, RLVR improved instruction following by 8--13 points (IFEval, IFBench) and complex reasoning by 5--6 points (OMEGA, ZebraLogic) across both model sizes, with detailed results available in the \MODEL technical report~\citep{olmo2025olmo3}.

\paragraph{Think models are far more expensive to post-train.} Excluding the shared thinking SFT, the Think-specific post-training (DPO + RLVR) requires ${\sim}$10K GPU hours for 7B and ${\sim}$97K GPU hours for 32B. The corresponding Instruct-specific post-training (instruction SFT + DPO + RLVR) requires ${\sim}$4.2K and ${\sim}$7.1K GPU hours, respectively. In datacenter energy terms, the gap is even wider: 32B Think post-training consumes 98,464 DC~kWh compared to 5,659 DC~kWh for 32B Instruct, a $17\times$ difference. This disparity is driven almost entirely by RLVR, where the Think models process far more tokens: the RLVR trainer processes $\sim$18.7B tokens for \MODEL 32B Think versus $\sim$564M for 32B Instruct ($33\times$), reflecting both the longer reasoning traces generated during rollouts and the greater number of training steps required.

\paragraph{The RLVR generator dominates.} Within RLVR, the generator (autoregressive rollout production) accounts for the vast majority of energy consumption: 87\% of all Think-specific post-training energy for \MODEL 32B, operating at an estimated ${\sim}$600W per GPU versus 130--220W for the trainer. The generator is functionally identical to large-scale inference deployment, providing real-world data on the energy cost of reasoning models. What makes RL training distinctively expensive is not the per-token cost (which is comparable to production inference) but the \emph{volume} and \emph{length} of generations required: our reasoning models generated rollouts averaging over 10,000 tokens, with a maximum of 32,768, repeated across many RL iterations. This total token volume is a hidden axis of cost that cannot be extrapolated from per-query inference costs alone.

The low trainer power draw (130--220W) reflects that trainer GPUs spend approximately 75\% of wall-clock time idle, waiting for rollouts from the generator. Despite significant engineering optimization, including continuous batching (which prevented an estimated 54\% waste from variable-length sequences), fully asynchronous training, and inflight weight updates, generation remained the overwhelming bottleneck, taking approximately $8\times$ longer than the corresponding training step. Even for the Instruct variant, the generator dominates: \MODEL 32B Instruct's RLVR generator consumes 3,575 DC~kWh versus 768 DC~kWh for the trainer ($4.7\times$), despite identical GPU hours. See Appendix~\ref{app:rlvr-details} for additional engineering details.

\subsection{Development costs}
\label{sec:dev-costs}

As in prior work~\citep{morrison2025holisticallyevaluatingenvironmentalimpact}, we report the environmental impact of model development: the hyperparameter searches, failed runs, data mixture experiments, and ablations that precede the final training runs reported in Table~\ref{tab:final-runs}. Figure~\ref{fig:pipeline} (bottom-left) breaks down total compute by training stage, separating final runs from development (see Table~\ref{tab:dev-costs} in the appendix for exact numbers).

Development accounts for 82.2\% of total GPU hours (6.85M of 8.34M) and 80.9\% of total GPU energy, excluding data generation. This is a substantial increase from the $\sim$50\% development fraction reported by \citet{morrison2025holisticallyevaluatingenvironmentalimpact} for pretraining. Reporting only the cost of final training runs would understate the true environmental impact by roughly $5\times$.

\paragraph{Post-training stages are especially iteration-heavy.} Pretraining has the lowest development fraction (68.5\%), because the final pretraining runs are large relative to the experiments that preceded them. All post-training stages, in contrast, have development fractions exceeding 91\%. Mid-training and data mixing show the highest development fraction at 97.5\%, reflecting extensive experimentation with data recipes, curricula, and domain mixtures. DPO is similarly high at 97.1\%, as the final DPO runs are small but many candidate approaches were evaluated. Even RL, which is itself expensive, has development fractions of 91--93\%, reflecting iteration on reward design, generation strategies, and hyperparameters.

\paragraph{Development costs grow with pipeline complexity.} The increase from $\sim$50\% to $\sim$82\% development cost is a direct consequence of adding more pipeline stages. Each stage (mid-training, SFT, DPO, RLVR) introduces its own hyperparameters, data choices, and design decisions that require iteration. A pretraining-only pipeline has one major development cycle; our multi-stage pipeline has at least six. As model development pipelines continue to grow in complexity, we expect the ratio of development to final run compute to increase further, making development cost reporting increasingly important for accurate environmental accounting.

\paragraph{Corroboration from financial data.} Our measurement-based finding is consistent with concurrent work by \citet{denain2026rnd}, who estimated from financial disclosures that final training runs account for only 10--23\% of total R\&D compute spending at three major AI companies (i.e., 77--90\% goes to development and experimentation). Our 82.2\% development fraction falls squarely within this range, despite using a fundamentally different methodology: we measure actual GPU hours and power consumption for a specific model series, while they estimate from company-wide financial data. They additionally note that R\&D workloads typically achieve lower hardware utilization than final training runs, which would make the gap even larger when measured in actual floating-point operations rather than dollars or GPU hours.

\section{Discussion}
\label{sec:discussion}

\subsection{The growing cost of post-training}
\label{sec:post-training-discussion}

As shown in Section~\ref{sec:cost-of-thinking}, post-training costs, particularly reinforcement learning, are becoming a substantial and under-reported component of the total environmental impact of building language models. Across our full pipeline, RL compute alone accounts for ${\sim}$1.4M GPU hours, or 17\% of total compute excluding offline data generation. For the Think variants, RL is the single most expensive post-training stage, and the RLVR generator dominates both GPU hours and energy consumption.

We expect this trend to accelerate for at least two reasons. First, reasoning models that produce long chains of thought are rapidly becoming the default paradigm, and longer outputs directly increase the compute required for rollout generation during RL. Second, the growing emphasis on tool use and agentic workflows, where models must generate sequences of actions, observations, and reasoning steps, will further amplify the token count per rollout, potentially by orders of magnitude compared to standard conversational queries. Both trends push the RL toward consuming an even larger share of compute.

Despite this trajectory, almost all environmental reporting in the field remains focused on pretraining. As post-training pipelines grow in complexity and cost, this gap in transparency will become increasingly consequential. We encourage model developers to report post-training costs alongside pretraining, broken down by stage where possible. \textbf{We strongly recommend that model developers report post-training costs alongside pretraining, broken down by stage, to make these growing costs visible.}

\subsection{Development costs grow with ambition}
\label{sec:dev-costs-discussion}

The increase in development cost fraction from $\sim$50\%~\citep{morrison2025holisticallyevaluatingenvironmentalimpact} to ${\sim}$82\% reflects a general pattern: as pipelines grow more complex, the ratio of experimentation to final training increases. Our development fraction is consistent with the 77--90\% estimated by \citet{denain2026rnd} for frontier-scale organizations, despite our models being substantially smaller, suggesting this is characteristic of modern LLM development in general.

The practical implication is that reporting only final training costs, as is standard practice, understates environmental impact by roughly $5\times$. This multiplier will likely increase as pipelines add more stages, each with its own experimentation cycle. The emergence of automated research frameworks that use LLMs to iterate on training recipes could further compound this effect, and the environmental cost of experimentation deserves the same scrutiny as the final training run. We note that this level of detailed accounting is feasible: we demonstrate it here for a complex multi-stage pipeline with multiple model variants, countering claims that such reporting is impractical. \textbf{At minimum, developers should report a development cost multiplier alongside final training costs.}

\subsection{Water: an infrastructure problem}
\label{sec:water}

Our analysis consumed an estimated 15,887~kL of water in total, equivalent to approximately 140 years of water consumption by the average person in the United States. While this figure is substantial, it is important to understand what drives it. Because our data center uses a closed-loop cooling system with no evaporative cooling ($WUE_{\text{onsite}} = 0$), \emph{all} of our water consumption comes from the power generation side ($WUE_{\text{offsite}}$): the water evaporated or consumed by thermo- and hydro-electric power plants. In our setup, water consumption is strictly proportional to energy consumption, and reducing one reduces the other.

This stands in contrast to data centers that use evaporative cooling towers, where $WUE_{\text{onsite}}$ can be substantial and water consumption scales with the amount of heat rejected, independent of the power source. Two data centers running the exact same AI workload can have very different water footprints depending entirely on their cooling infrastructure and the local power generation mix. This distinction is frequently lost in public discourse about AI's water impact, which tends to attribute water consumption to the AI workload rather than to the infrastructure choices made by datacenter operators and power providers.

We do not raise this point to dismiss the concern; water consumption from AI infrastructure is real and growing. Rather, our aim is to correctly attribute its causes. \textbf{Policy interventions aimed at reducing the water footprint of AI should focus on data center infrastructure standards (e.g., encouraging closed-loop or dry cooling) and power generation (e.g., shifting toward lower-water-intensity sources such as wind and solar)}, rather than on the characteristics of AI workloads themselves.

\subsection{Looking ahead}
\label{sec:looking-ahead}

We close with four concrete recommendations for improving environmental transparency in AI development:
\begin{enumerate}
    \item \textbf{Report full pipeline costs.} Post-training and data generation are substantial and growing. Developers should report costs for each stage, not just pretraining.
    \item \textbf{Include development costs.} Even a simple multiplier (e.g., ``development cost was $N\times$ the final training cost'') would be a significant improvement. We demonstrate here that detailed accounting of a complex multi-stage pipeline is feasible, countering the claim that such reporting is impractical.
    \item \textbf{Disclose datacenter efficiency metrics.} PUE, WUE, cooling type, and power source mix determine whether the same workload produces modest or severe environmental impact. These should be reported alongside compute costs.
    \item \textbf{GPU manufacturers should disclose manufacturing impacts.} Embodied emissions remain the least understood component of AI's environmental footprint, and will remain so until manufacturers provide real data.
\end{enumerate}
This work provides one piece of the puzzle: a detailed, measurement-based accounting of the environmental cost of building a modern multi-stage language model pipeline. We view this as a foundation for future work on data center siting, grid-level impacts of AI workloads, and policy recommendations for sustainable AI development.

\section*{Acknowledgments}

This material is based upon work supported by the National Science Foundation under Award Nos. 2413244 and 2326610, and by the National Science Foundation CISE Graduate Fellowships under Grant No. 2313998. Any opinions, findings, and conclusions or recommendations expressed in this material are those of the author(s) and do not necessarily reflect the views of the National Science Foundation.

\section*{Ethics statement}

This work aims to improve transparency around the environmental impact of AI development. We believe that accurate, detailed reporting of energy consumption, carbon emissions, and water usage is essential for informed decision-making by researchers, policymakers, and the public. Our environmental impact estimates rely on publicly available conversion factors and methodology; we note that these are approximations and that actual impacts may differ. Given that the scale and complexity of AI development will continue to grow, we believe realistic, measurement-based accounting is a necessary foundation for making informed decisions about sustainability in AI.

\bibliography{colm2026_conference}
\bibliographystyle{colm2026_conference}

\appendix

\section{Detailed per-run costs}
\label{app:detailed-runs}

\begin{table}[ht]
\begin{center}
\footnotesize
\begin{tabular}{llrrrr}
\toprule
& \textbf{Stage} & \textbf{GPU Hrs} & \textbf{DC kWh} & \textbf{CO$_2$ (kg)} & \textbf{Water (kL)} \\
\midrule
\multicolumn{6}{l}{\textit{Shared base model}} \\
\midrule
\multirow{4}{*}{7B}
  & Pretraining (6T) & 234,475 & 303,950 & 100,911 & 392 \\
  & Mid-training (100B) & 4,848 & 5,045 & 1,675 & 7 \\
  & Long context (50B) & 4,781 & 3,477 & 1,154 & 4 \\
  & Thinking SFT ($\sim$22.7B) & 2,409 & 2,894 & 961 & 4 \\
\cmidrule{2-6}
\multirow{6}{*}{32B}
  & Pretraining (6T) & 1,050,287 & 1,422,705 & 472,338 & 1,835 \\
  & Mid-training 1 (100B) & 14,549 & 20,306 & 6,742 & 26 \\
  & Mid-training 2 (100B) & 14,509 & 20,237 & 6,719 & 26 \\
  & Long context (50B) & 23,497 & 26,559 & 8,818 & 34 \\
  & Thinking SFT ($\sim$22.1B, lr=5e-5) & 9,015 & 12,417 & 4,123 & 16 \\
  & Thinking SFT ($\sim$22.1B, lr=8e-5) & 9,055 & 12,447 & 4,132 & 16 \\
\midrule
\multicolumn{6}{l}{\textit{Think-specific post-training}} \\
\midrule
\multirow{3}{*}{7B}
  & DPO & 286 & 190 & 63 & ${<}1$ \\
  & RLVR trainer ($\sim$8.7B) & 4,880 & 2,220 & 737 & 3 \\
  & RLVR generator & 4,880 & 6,114 & 2,030 & 8 \\
\cmidrule{2-6}
\multirow{3}{*}{32B}
  & DPO & 661 & 499 & 166 & 1 \\
  & RLVR trainer ($\sim$18.7B) & 27,400 & 12,148 & 4,033 & 16 \\
  & RLVR generator & 68,500 & 85,817 & 28,491 & 111 \\
\midrule
\multicolumn{6}{l}{\textit{Instruct-specific post-training}} \\
\midrule
\multirow{4}{*}{7B}
  & Instruction SFT ($\sim$1.7B) & 167 & 196 & 65 & ${<}1$ \\
  & DPO & 225 & 152 & 51 & ${<}1$ \\
  & RLVR trainer ($\sim$679M) & 472 & 157 & 52 & ${<}1$ \\
  & RLVR generator & 3,305 & 4,141 & 1,375 & 5 \\
\cmidrule{2-6}
\multirow{4}{*}{32B}
  & Instruction SFT ($\sim$1.7B) & 647 & 875 & 290 & 1 \\
  & DPO & 775 & 441 & 146 & 1 \\
  & RLVR trainer ($\sim$564M) & 2,853 & 768 & 255 & 1 \\
  & RLVR generator & 2,853 & 3,575 & 1,187 & 5 \\
\midrule
\multicolumn{2}{l}{\textbf{Total (final runs)}} & \textbf{1,485,329} & \textbf{1,947,327} & \textbf{646,513} & \textbf{2,512} \\
\bottomrule
\end{tabular}
\end{center}
\caption{Full per-stage cost breakdown for all \MODEL final training runs. The shared base model (pretraining through thinking SFT) is used by both Think and Instruct variants. For 32B, thinking SFT consists of two runs at different learning rates, followed by model merging. RLVR is split into a generator (inference, producing rollouts) and a trainer (optimization on rollouts). DC kWh is estimated total datacenter energy (GPU kWh $\times$ 1.74 server/IT overhead $\times$ 1.2 PUE). CO$_2$ uses $CI = 0.332$~kg/kWh; water assumes $WUE_{\text{offsite}} = 1.29$~L/kWh with $WUE_{\text{onsite}} = 0$ (closed-loop cooling).}
\label{tab:detailed-appendix}
\end{table}

\begin{table}[ht]
\begin{center}
\small
\begin{tabular}{lrrrr}
\toprule
\textbf{Stage} & \textbf{Total (GPU Hrs)} & \textbf{Final Run} & \textbf{Dev} & \textbf{Dev \%} \\
\midrule
Pretraining & 4,074,452 & 1,284,762 & 2,789,689 & 68.5\% \\
Mid-train \& data mix & 2,481,054 & 62,183 & 2,418,870 & 97.5\% \\
SFT & 299,635 & 21,293 & 278,342 & 92.9\% \\
DPO & 67,269 & 1,947 & 65,322 & 97.1\% \\
RL (trainer) & 503,049 & 35,605 & 467,444 & 92.9\% \\
RL (generator) & 909,581 & 79,537 & 830,044 & 91.3\% \\
\midrule
\textbf{Total} & \textbf{8,335,040} & \textbf{1,485,329} & \textbf{6,849,711} & \textbf{82.2\%} \\
\bottomrule
\end{tabular}
\end{center}
\caption{Development vs.\ final run compute by training stage. Development accounts for 82.2\% of total GPU hours. Post-training stages show particularly high development fractions ($>$91\%), reflecting extensive experimentation on data recipes, reward functions, and hyperparameters.}
\label{tab:dev-costs}
\end{table}

\section{RLVR engineering details}
\label{app:rlvr-details}

As discussed in Section~\ref{sec:cost-of-thinking}, the RLVR trainer spends approximately 75\% of wall-clock time idle, waiting for rollouts. Variance in generation length means straggling completions (e.g., very long reasoning chains approaching the 32K token limit) gate effective throughput, and simply allocating more generator GPUs does not fully resolve this. Active sampling strategies that prioritize high-signal prompts can reduce total generation volume, but the fundamental asymmetry between generation and optimization remains.

\paragraph{Power profiles across stages.} The average per-GPU power draw differs substantially across training stages. Pretraining sustains the highest utilization at 620--649W, as computation and communication are well-overlapped. SFT varies widely (560--660W) depending on sequence length and batch configuration. DPO runs at moderate power (272--361W). The starkest contrast is within RLVR: the generator operates at near-peak power (${\sim}$600W), while the trainer runs at a fraction of that (130--220W), limited by the rate at which new rollouts arrive.

\end{document}